\documentclass[fleqn,usenatbib]{mnras}

\usepackage{newtxtext,newtxmath}

\usepackage[T1]{fontenc}

\DeclareRobustCommand{\VAN}[3]{#2}
\let\VANthebibliography\thebibliography
\def\thebibliography{\DeclareRobustCommand{\VAN}[3]{##3}\VANthebibliography}

\usepackage{graphicx}	
\usepackage{amsmath}	
\usepackage{float}
\usepackage{geometry}
\usepackage{booktabs}
\usepackage{lscape}
\usepackage{threeparttable}
\usepackage{xcolor}
\usepackage{ulem}
\usepackage{cancel}
\usepackage{subcaption}

\newcommand{\nickel}{\ensuremath{^{56}\mathrm{Ni}}}

\newcommand{\mch}{\ensuremath{M_\mathrm{Ch}}}
\newcommand{\artis}{\textsc{artis}}
\newcommand{\tilda}{$\sim $ }
\newcommand{\msun}{\ensuremath{\mathrm{M}_\odot}}

\title[Helium in double detonation models]{Exploring the range of impacts of helium in the spectra of double detonation models for Type Ia supernovae}

\author[F. P. Callan et al.]{F. P. Callan,$^{1}$\thanks{E-mail: f.callan@qub.ac.uk}
C. E. Collins,$^2$ S. A. Sim,$^{1}$ L. J. Shingles,$^2$ R. Pakmor,$^3$ S. Srivastav,$^4$ J. M. Pollin,$^1$ \newauthor S. Gronow,$^5$ F. K. R\"opke$^{5,6,7}$ and I. R. Seitenzahl$^5$
\\
$^{1}$School of Mathematics and Physics, Queen's University Belfast, University Road, Belfast BT7 1NN, UK \\
$^{2}$GSI Helmholtzzentrum f\"{u}r Schwerionenforschung, Planckstraße 1, 64291 Darmstadt, Germany\\
$^3$Max-Planck-Institut f\"{u}r Astrophysik, Karl-Schwarzschild-Str. 1, D-85748, Garching, Germany\\
$^4$Astrophysics sub-Department, Department of Physics, University of Oxford, Keble Road, Oxford, OX1 3RH, UK\\
$^5$Heidelberger Institut f\"ur Theoretische Studien, Schloss-Wolfsbrunnenweg 35, D-69118, Heidelberg, Germany\\
$^6$Zentrum f{\"u}r Astronomie der Universit{\"a}t Heidelberg, Institut f{\"u}r Theoretische Astrophysik, Philosophenweg 12, D-69120 Heidelberg, Germany\\
$^7$Zentrum f{\"u}r Astronomie der Universit{\"a}t Heidelberg, Astronomisches Rechen-Institut, M{\"o}nchhofstr, 12--14, 69120 Heidelberg, Germany\\
}

\date{Accepted XXX. Received YYY; in original form ZZZ}

\pubyear{2024}

\begin{document}
\label{firstpage}
\pagerange{\pageref{firstpage}--\pageref{lastpage}}
\maketitle

\begin{abstract}
Models of sub-Chandrasekhar mass double detonations for Type Ia supernovae (SNe~Ia) suggest a distinguishing property of this scenario is unburnt helium in the outer ejecta. However, modern explosion simulations suggest there may be significant variations in its mass and velocity distribution. We recently presented a NLTE (non local thermodynamic equilibrium) radiative transfer simulation for one realisation of the double detonation scenario with a modest He mass (0.018\,\msun) present in the ejecta at relatively high velocities (${\sim}18000\,\mathrm{km}\,\mathrm{s}^{-1}$). That simulation predicted a \ion{He}{I}\,10830 \AA\ feature blueward of \ion{Mg}{II} 10927\,\AA\ consistent with near-infrared observations of ‘‘transitional'' SNe Ia. To demonstrate the expected diversity in the helium signature, here we present a calculation for a double detonation model with a higher He mass (${\sim}$0.04\,\msun) ejected at lower velocities (${\sim}13000\,\mathrm{km}\,\mathrm{s}^{-1}$). Despite our simulation predicting no clear optical or 2 $\mu$m helium features, a strong and persistent \ion{He}{I}\,10830~\AA~absorption is present. The feature appears at wavelengths consistent with the extended blue wing of the \ion{Mg}{II} 10927\,\AA\ feature sometimes present in observations, suggesting this is a helium spectral signature (although for this particular model it is too strong and persistent to be consistent with normal SNe Ia). The significant differences in \ion{He}{I}\,10830~\AA\ predicted by the two simulations suggests helium spectral signatures likely show significant variation throughout the SNe Ia population. This motivates further work to use this observable signature to test the parameter space for double detonation models.
\end{abstract}




\section{Introduction}
\label{sec:intro}
Type Ia supernovae (SNe Ia) are the thermonuclear explosions of at least one white dwarf (WD) in a close binary system. However, the nature of the progenitor systems and explosion mechanisms that produce SNe~Ia are still uncertain. The {\it double detonation} is a widely discussed sub-Chandrasekhar (sub-\mch) explosion mechanism which has the potential to explain SNe~Ia with a variety of luminosities. This scenario involves the ignition of a He detonation in a surface He layer (shell) on a carbon-oxygen (CO) WD \citep[see, e.g., the study of He ignition by][]{glasner2018a}. The shock from the initial He detonation then results in the ignition of a secondary core detonation that entirely disrupts the WD. The mass of He required for a He detonation to occur depends on the specific explosion mechanism. For spontaneous ignition following stable mass transfer from a He star, massive He shells (> $0.1$\,\msun) are required (e.g.\,\citealt{nomoto1980a, nomoto1982a, taam1980a, livne1990a, woosley1994a, hoeflich1996a, nugent1997a}). This results in light curves and spectra incompatible with observations of normal SNe~Ia due to the over-abundance of iron-group elements in the outer ejecta produced by the He detonation. If the stable mass transfer is instead from a degenerate companion, significantly less massive He shells are required (< $0.1$\,\msun\ \citealt{bildsten2007a, shen2009a, shen2010a}). We note that while it has been demonstrated that if such a He detonation occurs, a secondary core detonation will likely follow \citep{fink2010a, gronow2020a, gronow2021a, boos2021a}, the initiation of the He detonation may be inhibited when rapid rotation is considered (see e.g.\,\citealt{neunteufel2017a, neunteufel2019a, piersanti2024a}). In an alternative scenario, the He detonation may be triggered dynamically during the merger of two WDs \citep{shen2014a, tanikawa2018a, tanikawa2019a, pakmor2022a, shen2024a}, requiring only low He shell masses (< 0.1\,\msun). Double detonation simulations invoking low He shell masses predict light curves and spectra that are much more consistent with normal SNe~Ia (e.g. \citealt{kromer2010a, woosley2011a, polin2019a, shen2021a, pollin2024a, collins2025a} although see also \citealt{collins2022a}).

\cite{noebauer2017a} predict that the radioactive material produced in the outer layers of the ejecta from He shell burning in double detonation models can result in early light curve flashes. Additionally, simulations of double detonation models with more massive He shells ($\gtrsim$0.05\,\msun, e.g.\,\citealt{kromer2010a, sim2012a, polin2019a, collins2022a}) predict significant amounts of ultra-violet (UV) line blanketing due to the iron group elements synthesised in the He shell burning. A number of observed SNe~Ia which exhibit either or both of these signatures have already been suggested as double detonation candidates (e.g.\,\citealt{inserra2015a, jiang2017a, de2019a, jacobson-galan2020a, dong2022a, ni2022a, liu2023a, padilla-gonzalez2023a, padilla-gonzalez2024a}). 

Another key characteristic of double detonation explosion models is that they predict significant masses of unburnt He in the ejecta (${\sim}0.01{-}0.1$\,\msun, see e.g.\,\citealt{fink2010a, shen2014a, gronow2020a, gronow2021a, pakmor2022a}). Here we focus on the direct detection of this unburnt He as a signature of the double detonation scenario. The He ionization and excitation rates in supernova ejecta are strongly impacted by collisions with non-thermal electrons produced by $\gamma$ rays from \nickel~decays. NLTE (non local thermodynamic equilibrium) radiative transfer simulations that include a treatment for non-thermal processes are therefore required to predict if He spectral features are expected to form. As such, many previous radiative transfer simulations (e.g.\,\citealt{kromer2010a, woosley2011a, sim2012a, polin2019a, townsley2019a, gronow2020a, shen2021a, collins2022a, pakmor2022a, pollin2024a}) have been unable to predict if He spectral features form. 

The importance of collisions with non-thermal electrons for the formation of He features in type Ib/Ic supernovae has already been demonstrated (e.g.\,\citealt{chugai1987a, lucy1991a, hachinger2012a}). The formation of \ion{He}{I} spectral features has also been investigated in previous radiative transfer simulations of SNe~Ia models that include treatment for non-thermal electron collisions. \cite{nugent1997a} carried out a radiative transfer simulation for the thick He shell (0.2\,\msun) double detonation model of \cite{woosley1994a} and reported no clear signatures of He in the model spectra, although the \ion{He}{I}\,10830\,\AA~line did show substantial optical depths at some points in the model. Additionally, as part of their investigation of the prominent spectral absorption feature observed in the near-infrared (NIR) spectra of SN~1994D around 10500\,\AA, which \citet{meikle1996a} suggest could be either \ion{He}{I}\,10830 \AA~or \ion{Mg}{II} 10927\,\AA, \cite{mazzali1998a} found that by introducing a few hundredths of a solar mass of He into the W7 deflagration model of \cite{nomoto1984a} a \ion{He}{I}\,10830\,\AA~spectral feature could form. \cite{dessart2015a} also showed that He spectral features formed in their radiative transfer simulations of a He detonation model by \cite{waldman2011a}. However, this model only considers a He detonation and produces light curves that are able to match only faint and fast declining events.

More recently \cite{boyle2017a} investigated whether He spectral features were expected to form for a double detonation model with a modest ejected He mass of 0.03\,\msun~(based on the explosion model of \citealt{fink2010a}). By using an analytical approximation to estimate the excited \ion{He}{i} level populations they found \ion{He}{I}\,10830 \AA~and 2 $\mu$m lines may be formed by high velocity (${\sim}19000\,\mathrm{km}\,\mathrm{s}^{-1}$) He and the \ion{He}{I}\,10830 \AA~feature predicted  may be consistent with a prominent absorption feature observed around 10300\,\AA~in SN 2016dsg, an observed double detonation candidate \citep{dong2022a}. However, \cite{boyle2017a} utilised an approximate NLTE treatment that assumes the \ion{He}{II} population is dominant. As it is not clear whether the non-thermal ionization rate remains high enough to ensure He remains ionized, only an upper limit on the potential strength of the He features is possible with this approach. Therefore, radiative transfer simulations with a self-consistent calculation of the ionization state are required to investigate if He spectral features are expected for double detonation models that adopt more modest He shell masses. \cite{collins2023a} [hereafter C23] carried out such a simulation for the M2a double detonation model of \cite{gronow2020a} utilising a full NLTE treatment of the plasma conditions and also accounting for non-thermal electrons. This simulation showed a clear high velocity (${\sim}19000\,\mathrm{km}\,\mathrm{s}^{-1}$) \ion{He}{I}\,10830 \AA~feature that matches the evolution of a feature in the spectrum of the transitional SN~Ia iPTF13ebh \citep{hsiao2015a}, demonstrating the potential of He spectral features as a signature of the double detonation scenario. However, due to blending with strong lines the exact contribution of \ion{He}{I} to the optical spectra in the C23 model is not clear. Additionally, as a result of the limited signal to noise of the simulation, it was not possible to confirm whether the He 2 $\mu$m spectral feature, which could act as an independent confirmation of the presence of He, was expected to form in the model. 

The amount of unburnt He, and its range of ejection velocity in the double detonation scenario can show significant variation for different lines of sight  and between models (see e.g.\,figure\,1 in \citealt{collins2022a}). Explosion simulations for models in which the He detonation is assumed to be ignited through compressional heating of accreted material predict very little unburnt He is present at velocities below ${\sim}10000\,\mathrm{km}\,\mathrm{s}^{-1}$. In another widely discussed double-detonation scenario, the initial He shell detonation is instead ignited during the dynamical merger of two CO WDs which leads to a core detonation of the primary WD \citep{guillochon2010a, pakmor2013a, pakmor2022a, tanikawa2018a, tanikawa2019a, boos2021a, boos2024a, shen2021b, pollin2024a}. In this scenario the secondary WD may also explode via a double detonation \citep{tanikawa2019a, pakmor2022a, boos2024a}. Explosion simulations of such models predict unburnt He at generally lower velocities compared to accretion ignited double detonation models and for some lines of sight also show significant amounts of unburnt He at velocities below ${\sim}10000\,\mathrm{km}\,\mathrm{s}^{-1}$ (see e.g. figure\,1 of \citealt{pollin2024a}). The model presented by C23 is representative of a realisation of the double detonation scenario with He present at relatively high velocities (the He distribution peaks at ${\sim}18000\,\mathrm{km}\,\mathrm{s}^{-1}$) and a reasonably large mass of \nickel~present in the He shell detonation ash. In this paper, we instead investigate the potential of He spectral features to form for a realisation of the double detonation scenario with unburnt He present at significantly lower velocities (the He distribution peaks at ${\sim}13000\,\mathrm{km}\,\mathrm{s}^{-1}$) and only a small mass of \nickel~present in the He shell detonation ash. In particular, comparing the He spectral features predicted by our new model to those predicted by the model of C23 will provide insight into how sensitive the spectral signatures of He are to different model ejecta structures. 

Throughout the study we compare the synthetic spectra produced to the normal SN 2011fe \citep{nugent2011a} and the ‘‘transitional'' SN 2022xkq \citep{pearson2024a}, which showed a similar feature around 10300\,\AA~to the feature C23 identified as \ion{He}{I}\,10830\,\AA~in the spectrum of SN~iPTF13ebh. Note, the ‘‘transitional'' SNe~Ia we refer to throughout this work belong to subclass of fast declining SNe~Ia described by \cite{hsiao2015a} that photometricaly bridge the gap between subluminous 91bg-like and normal SNe~Ia.

We describe our model ejecta structure and the radiative transfer simulation set up in Section\,\ref{sec:numerical_methods}. We then present our simulated spectra and comparisons with those of C23 as well as SN 2011fe and SN 2022xkq in Section\,\ref{sec:results} before presenting our conclusions in Section\,\ref{sec:conclusions}.

\section{Numerical Methods}  
\label{sec:numerical_methods}

\subsection{NLTE and non-thermal radiative transfer}
\label{subsec:RT_method}
The radiative transfer simulations are carried out using the time-dependent multi-dimensional Monte Carlo radiative transfer code \artis\footnote{\href{https://github.com/artis-mcrt/artis/}{https://github.com/artis-mcrt/artis/}}~\citep{sim2007b, kromer2009a, bulla2015a, shingles2020a}. \textsc{artis} follows the methods of \citet{lucy2002a, lucy2003a, lucy2005a} and is based on dividing the radiation field into indivisible energy packet Monte Carlo quanta. In this work (as in C23) we utilise the full NLTE and non-thermal capabilities added to \artis~by \citet{shingles2020a}. This includes a NLTE population and ionization solver and treatment for collisions with non-thermal leptons. To follow the energy distribution of high energy leptons which result from nuclear decays and Compton scattering of $\gamma$-rays, \artis~solves the Spencer-Fano equation (as framed by \citealt{kozma1992a}). Auger electrons are allowed to contribute to heating, ionization and excitation. Excitation of bound electrons by non-thermal collisions is also included which was not the case for the simulation of C23. 

For photoionization, following \cite{lucy2003a}, \artis~utilises the full Monte Carlo photon-packet trajectories to obtain a rate estimator for each level pair while a parameterized radiation field is used to estimate transition rates for all other atomic processes. The atomic data set we use is based on the compilation of \textsc{cmfgen} \citep{hillier1990a, hillier1998a} and is similar to that described by \cite{shingles2020a} but with some additional ions. In this simulation we include He \textsc{i-iii}, C \textsc{i-iv}, O \textsc{i-iv}, Ne \textsc{i-iii}, Mg \textsc{i-iii}, Si \textsc{i-iv}, S \textsc{i-iv}, Ar \textsc{i-iv}, Ca \textsc{i-iv}, Ti \textsc{ii-iv}, Cr \textsc{i-v}, Fe \textsc{i-v}, Co \textsc{ii-v} and Ni \textsc{ii-v}. 

The simulations also utilise the virtual packet scheme \cite{bulla2015a} implemented into \artis~(which C23 did not utilise). Virtual packets significantly improve the signal to noise of the synthetic spectra produced with only a small computational penalty. Moreover, the virtual packet approach (see \citealt{bulla2015a} for details) enables additional spectra to be extracted, in which the line opacity of specified atomic species are neglected, allowing their spectral contribution to be isolated. This is particularly useful for our investigation of potential NIR and optical He spectral features. 

In previous work, C23 demonstrated that for the M2a double detonation model of \cite{gronow2020a} the spectral contribution of He is strongest in the first few days after explosion and then becomes weaker with time until it is no longer clearly contributing by two weeks after explosion. We therefore focus on relative early times in this work: the radiative transfer simulations are carried out between 1.5 and 35~d relative to explosion explosion, using 115 logarithmically spaced time steps. We adopt a grey approximation in optically thick cells (those with a Thomson optical depth greater than 1000). The initial time steps are treated in local thermodynamic equilibrium (LTE) and at time step 12 (2.1\,d) the NLTE solver is switched on.  

\subsection{Double detonation ejecta model}
\label{subsec:ejecta_model}

\begin{figure*}
	\includegraphics[width=0.9\linewidth,trim={1.5cm 0cm 2cm 1cm},clip]{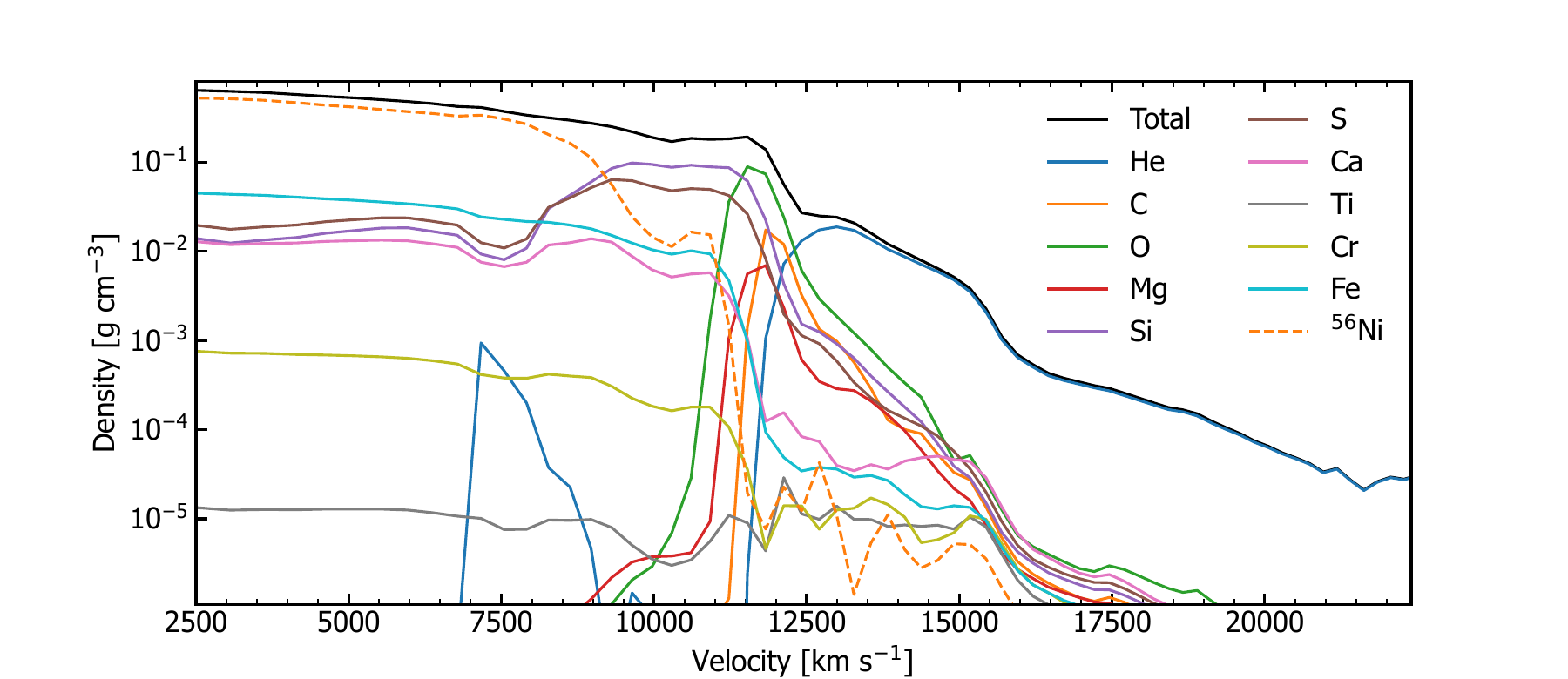}    
    \caption{Model ejecta composition at 100\,s after explosion.}
    \label{fig:Ejecta_composition}
\end{figure*}

The ejecta structure we simulate here is drawn from the 3D double detonation explosion model M08\_03, described by \cite{gronow2021a}. While \artis~has the ability to carry out multi-dimensional simulations, utilising the full NLTE and non-thermal capabilities of \artis~required to accurately assess whether He spectral features form is computationally expensive and has significant memory requirements owing to the large number of Monte Carlo estimators involved. Therefore, for this work we construct a 1D model from the 3D M08\_03 ejecta model. We note that the ejecta of SNe~Ia explosion models are inherently multi-dimensional (e.g.\,\citealt{pakmor2024a}). In particular we note there are angular variations in the mass fractions of \nickel, He and the velocities at which material is ejected (see \citealt{collins2022a} figure 1). These have significant impact as demonstrated by the approximate NLTE 3D radiative transfer simulation of the M08\_03 model by \cite{collins2022a} that showed synthetic spectra with strong viewing angle dependencies. 
However, the 1D simulation presented here is an important initial investigation of the parameter space of potential He spectral signatures and will help identify which models may be of most interest to explore with multi-dimensional simulations in future work. 

When constructing our 1D model we radially average cells in the 3D model that are contained within a solid angle cone centred on the negative $z$-axis to attempt to preserve the representative ejecta stratification and density profile of this radial direction in the 3D model. This is particularly important for investigating double detonation models because the velocity of the He shell ash varies noticeably with polar angle in the 3D double detonation models of \cite{gronow2021a} [see figure\,1 in \citealt{collins2022a}]; thus if we impose spherical averaging we would effectively be artificially smoothing out the boundary region between the He shell ash and the core ejecta (in the 3D models, this boundary region is fairly sharply defined, but occurs at different velocities as a function of polar angle). We neglect the low density outer ejecta ($v$ > 26200 km s$^{-1}$) that become quickly optically thin and thus are not expected to contribute to the spectral formation for the times considered here. The ejecta composition of our model is shown in Figure~\ref{fig:Ejecta_composition}.   

The 1D ejecta model presented here has a total mass of 1.01\,\msun, a He mass of 0.04\,\msun~and a \nickel~mass of 0.47\,\msun whereas the original 3D ejecta model has a total mass of 0.83\,\msun, a He mass of 0.018\,\msun~and a \nickel~mass of 0.13\,\msun. The greater He, \nickel~and total mass of the 1D model are a result of significant asymmetries in the mass distribution of the 3D model with more mass ejected towards the negative $z$-axis (again see figure\,1 in \citealt{collins2022a}). Thus, the 1D model constructed here is not a good representation of the original 3D model as a whole and will not preserve either its total energetics or luminosity. However, it does preserve the realistic radial structure from a representative double-detonation model, and has masses that are within the range expected for double detonation models (see e.g.\,table 3 in \citealt{fink2010a}). We note the greater mass of both \nickel~and He in our model mean this model is likely representative of a line-of-sight exhibiting generally stronger He spectral features than the majority of the model viewing angles due to the increased number of non-thermal collisions with He expected. 

The ejecta model simulated in this study is based on the negative $z$ (south pole) region of the 3D M08\_03 explosion model \cite{gronow2020a} presented (the He detonation is ignited on the positive $z$ axis) while the C23 model is based on an equatorial slice of the 3D M2a explosion model presented by \cite{gronow2020a}. As such the models show significant differences in their ejecta: although our model has a \nickel~mass similar to the C23 ejecta model (0.47 and 0.49 \msun~respectively) it has a total He mass (0.04 \msun) which is more than double that of the C23 model (0.018\,\msun). Additionally, the ejecta structures show significant differences (compare Figure~\ref{fig:Ejecta_composition} and figure 1 in C23). In particular, for the C23 model the He lies at a relatively high velocity (peaking at ${\sim}18000\,\mathrm{km}\,\mathrm{s}^{-1}$) and there is a significant mass of \nickel~($10^{-2}$\,\msun) present in the He shell. In contrast, the ejecta model investigated in this study shows a substantial mass of unburnt He at significantly lower velocities (peaking at ${\sim}13000\,\mathrm{km}\,\mathrm{s}^{-1}$) with only a very small mass of \nickel~(<$10^{-4}$\,\msun) present in this region (such low \nickel~to He ratios are characteristic of the negative $z$ region in the \cite{gronow2021a} models, see \citealt{collins2022a} figure 1). As a result our new model has significantly less \nickel~co-spatial with the unburnt He in the He shell relative to the C23 model. Therefore, the radiative transfer simulation presented here allows us to explore a significantly different realisation of the double detonation scenario than C23. 

\section{Results}
\label{sec:results}

\subsection{Helium spectral features}
\label{subsec:He_features}
\subsubsection{NIR \ion{He}{I} 10830 \AA}
\label{subsubsec:He_I_10830}
\begin{figure}
    \centering
    \begin{subfigure}[t]{0.1667\textwidth}
        \centering
	\includegraphics[width=1\linewidth,trim={0.0cm 2.0cm 0.8cm 4.2cm},clip]{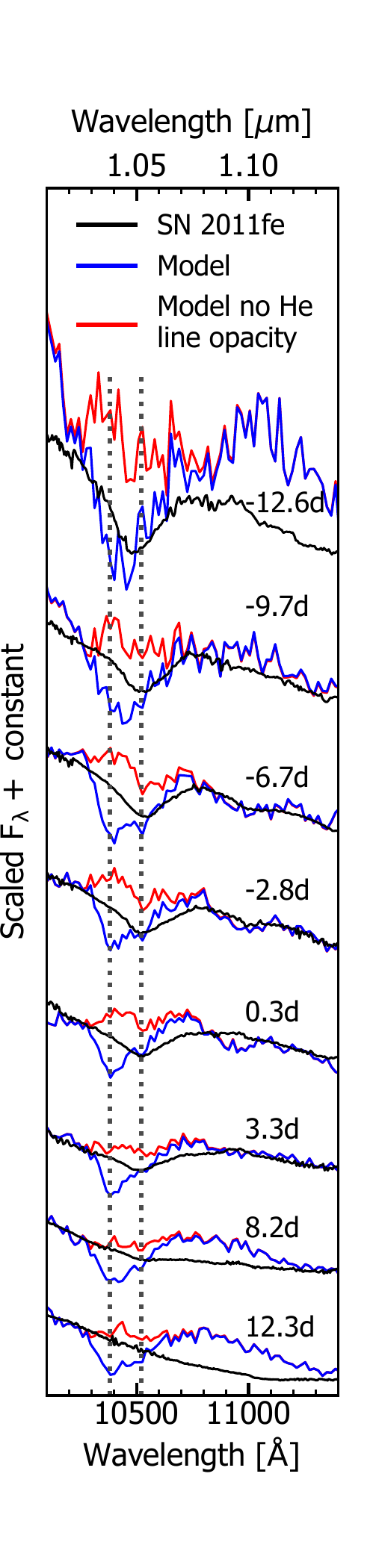} 
    \end{subfigure}
    \hfill
    \begin{subfigure}[t]{0.145\textwidth}
        \centering

    \includegraphics[width=1\linewidth,trim={1cm 2.0cm 0.8cm 4.2cm},clip]{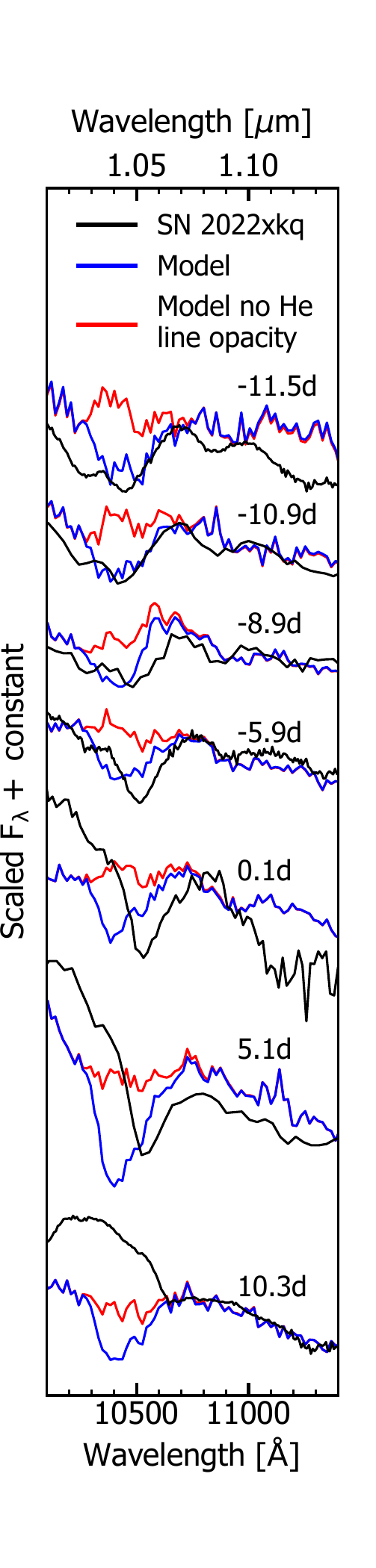} 
    \end{subfigure}
    \hfill
    \begin{subfigure}[t]{0.145\textwidth}
        \centering
	\includegraphics[width=1\linewidth,trim={1cm 2.0cm 0.8cm 4.2cm},clip]{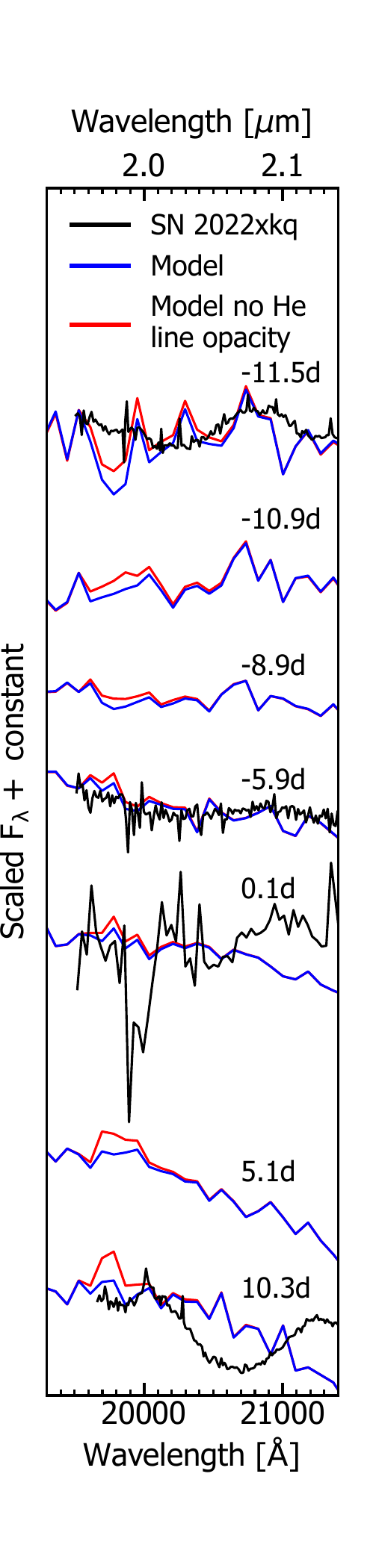}    
    \end{subfigure}
    \caption{Evolution of model \ion{He}{I}\,10830 \AA~feature compared to SN 2011fe \citep{hsiao2013a} and SN 2022xkq \citep{pearson2024a}. Also shown is the evolution of the \ion{He}{I} 2 $\mu$m contribution predicted by the model compared to SN 2022xkq. In each case the full model spectra and the model spectra in which He line opacity is omitted are plotted.   Times are relative to \textit{B}-peak. The deepest absorption at 0.3\,d for the simulated \ion{He}{I}\,10830 \AA~feature (left) and the \ion{Mg}{II} 10927\,\AA~feature observed for SN 2011fe (right) are indicated by the dotted lines for reference. Note, as the spectrum of SN 2022xkq is impacted by a telluric region redward of 2 $\mu$m this region of the spectrum of SN 2022xkq is not plotted in the right panel of the Figure for clarity.}
    \label{fig:NIR_features_zoom}
\end{figure}

By far the strongest He spectral feature in our simulation is \ion{He}{I} 10830~\AA. Figure\,\ref{fig:NIR_features_zoom} shows the NIR spectroscopic evolution of our model around this feature (see also Figure~\ref{fig:NIR_spectral_series}). We also show the same calculation when the line opacity of He is omitted (as discussed in Section \ref{subsec:RT_method}) to illustrate the impact on the spectrum of He absorption. 

The model predicts a strong \ion{He}{I}\,10830 \AA~feature that persists for all epochs simulated. Although the profile shape changes with time the absorption is consistently around ${\sim}12000\,\mathrm{km}\,\mathrm{s}^{-1}$. This is close to the velocity at which the density of unburnt He peaks (see Figure \ref{fig:Ejecta_composition}). The strength of this feature in our simulation confirms that \ion{He}{i} 10830 \AA~can be an observable signature of the double detonation explosion scenario which provides a potentially powerful discriminant between double detonation models and other explosion scenarios which do not eject significant amounts of He. 

\subsubsection{Optical \ion{He}{i}}
\label{subsubsec:He_Optical}
Ideally, to confirm the presence of He in observations, we would like to identify a secondary He feature to corroborate the identification of \ion{He}{i} 10830\,\AA. 
We therefore discuss the optical He features predicted by our simulation and the potential for them to be observed.

\begin{figure*}
    \centering
    \begin{subfigure}[t]{0.495\textwidth}
        \centering
  \includegraphics[width=1\linewidth,trim={2.7cm 3.8cm 3.0cm 6.0cm},clip]{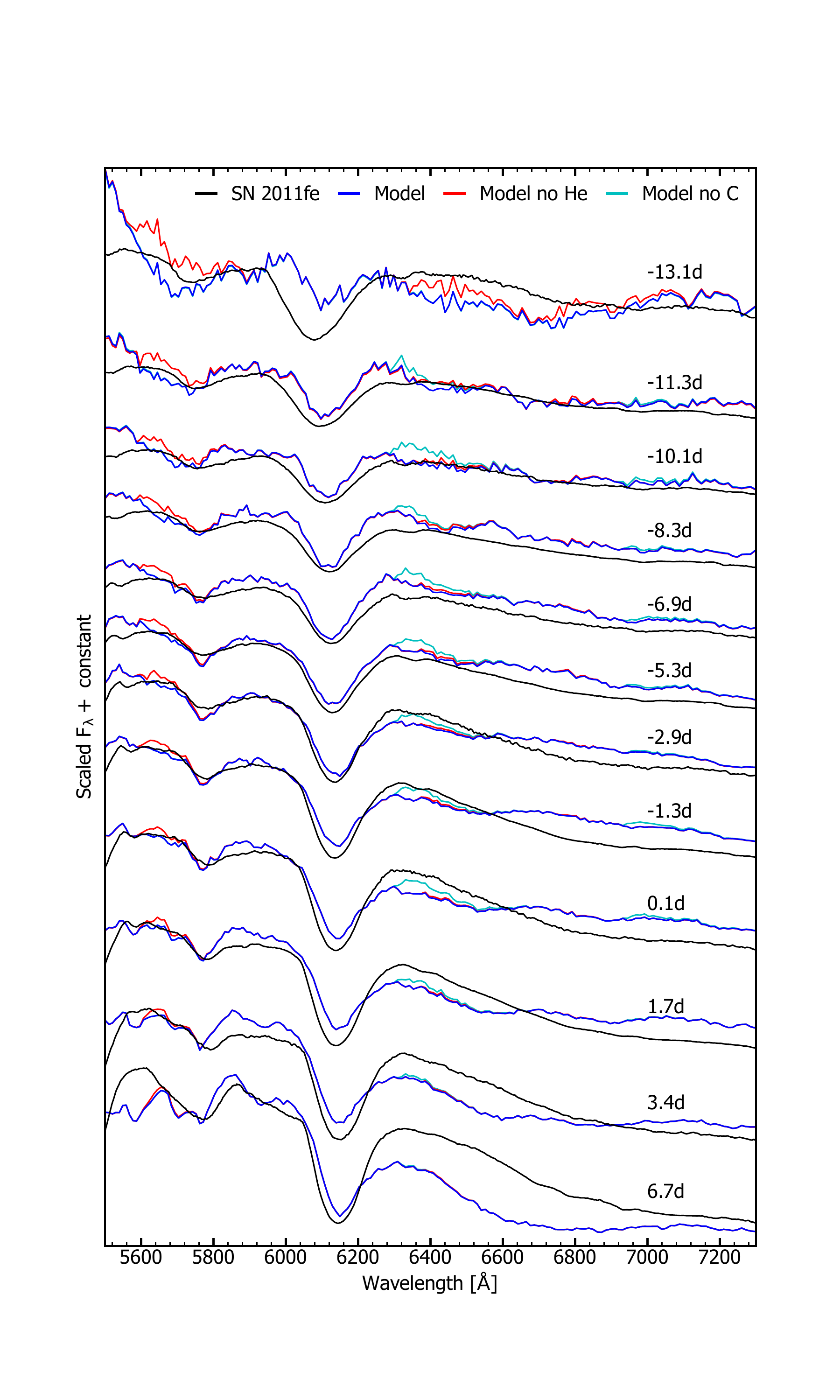}    
    \end{subfigure}
    \hfill
    \begin{subfigure}[t]{0.495\textwidth}
        \centering
 \includegraphics[width=1\linewidth,trim={2.7cm 3.8cm 3.0cm 6.0cm},clip]{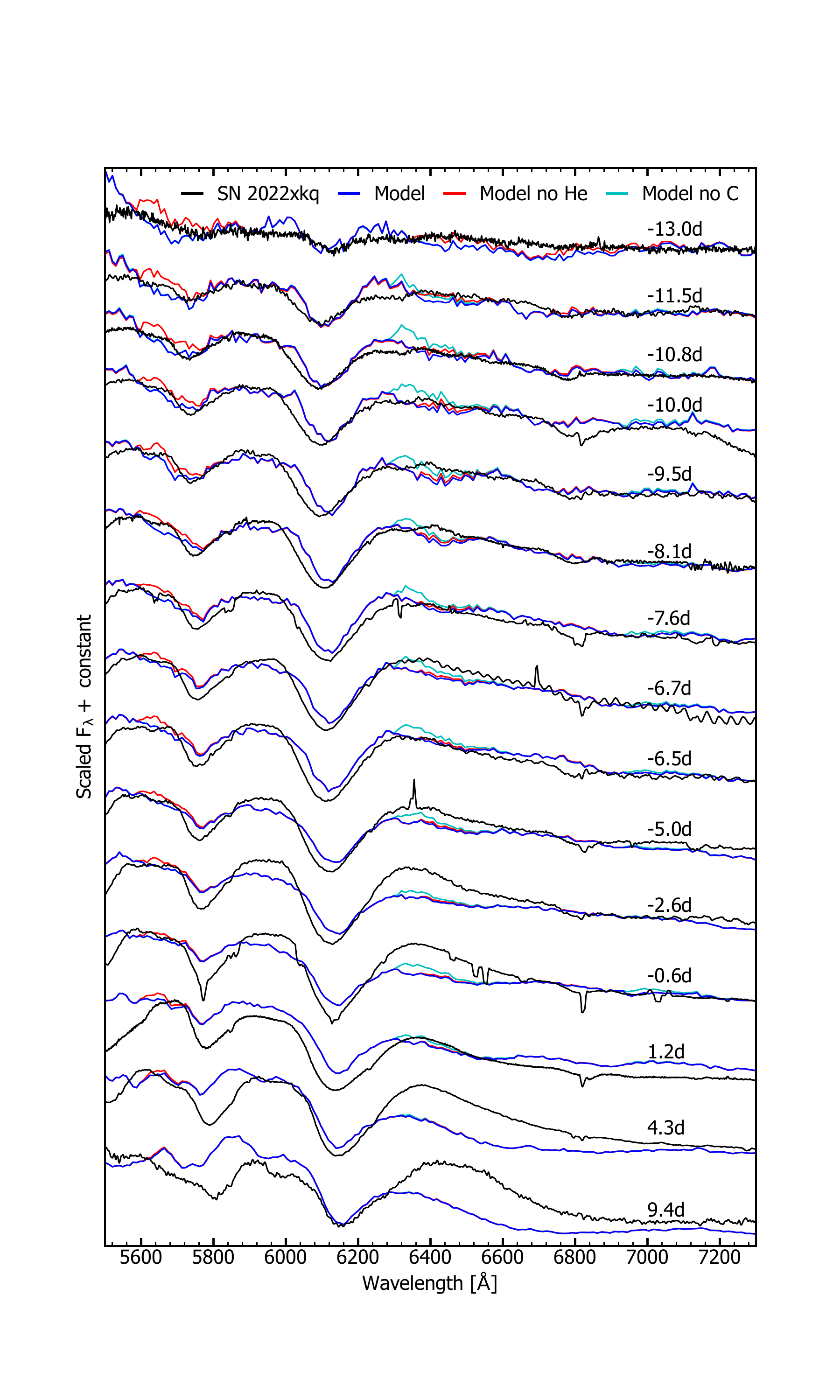}    
    \end{subfigure}
    \caption{Model spectroscopic evolution of optical \ion{He}{I} (5876 and 6678 \AA) and \ion{C}{II} (6580 and 7235 \AA) lines compared to that of SN 2011fe \citep{pereira2013a, mazzali2014a} and SN 2022xkq \citep{pearson2024a}. Also plotted are the model spectra in which the line opacities of He and C are not considered. Times are relative to \textit{B}-peak.}
    \label{fig:Optical_He_spectral_series}
\end{figure*}

The He line that shows the strongest contribution to the optical spectra of our model (as quantified by the difference between the full spectra and the spectra lacking He line opacity) is \ion{He}{i} 5876\,\AA. Some packets in the simulation do interact with the \ion{He}{i} 5876\,\AA~line for all epochs shown in Figure \ref{fig:Optical_He_spectral_series}, with the simulation predicting the number of interactions is greatest at early times and decreases with time. However, this does not result in a clear spectral feature forming in the model with the contribution of \ion{He}{i} 5876\,\AA~instead blending with the blue wing of the \ion{Si}{ii} 5972\,\AA~feature. The next most prominent contribution of He to our model optical spectra is from \ion{He}{i} 6678\,\AA. Again, while some packets are interacting with the \ion{He}{i} 6678\,\AA~line (until around \textit{B}-peak) no clear spectral feature associated with this line is predicted by our model.  

Thus, while this simulation predicts a strong and persistent \ion{He}{i} 10830\,\AA~feature, we do not find clear optical He features. This confirms that a lack of identifiable optical He features can not be used to rule out the presence of He, as previously demonstrated by the radiative transfer simulations of \cite{mazzali1998a} and C23. We also note \cite{nugent1997a} found no strong evidence for optical He in their radiative transfer simulation of a double detonation model. 

\subsubsection{NIR \ion{He}{i} 2 $\mu$m}
\label{subsubsec:2micron_He}

Since optical He features can not be used to corroborate the \ion{He}{I} 10830\,\AA~feature identification in observations, we now discuss the possibility of identifying a \ion{He}{i} $2 \mu$m feature in our simulation. Comparing the full model spectra to the model spectra in which He line opacity is neglected in Figure~\ref{fig:NIR_features_zoom} (and Figure~\ref{fig:NIR_spectral_series}) we see that only a small number of packets in our simulation are interacting with the \ion{He}{I} 2 $\mu$m line. We note that the spectral contribution of the \ion{He}{I} 2 $\mu$m is greatest for the latest two epochs shown (5.1 and 10.3\,d after \textit{B}-peak), although it is still minor. Thus our simulation does not predict any clear \ion{He}{I} 2 $\mu$m spectral feature is formed at any epochs considered here.  

\subsection{Comparisons to C23 and implications for He signature diversity}
\label{subsec:C23_comparisons}
As discussed in Section \ref{subsec:ejecta_model} the model presented here represents a double detonation structure with a relatively substantial mass of ejected He (0.04 \msun), a significant fraction of He present at reasonably low velocities and only a small mass of \nickel~present in the He shell detonation ash. In contrast the double detonation model presented by C23 has a significantly lower mass of ejected He (0.018\,\msun), He present at higher velocities and a reasonably large mass of \nickel~present in the He shell detonation ash. In the following we discuss the differences in the He features predicted by each of these models and comment on the implications of these differences for the double detonation scenario.

The \ion{He}{I}\,10830 \AA~feature predicted by our model forms at an approximately constant velocity (${\sim}12000\,\mathrm{km}\,\mathrm{s}^{-1}$) for all epochs simulated. The inner velocity of the He rich region in our model is at ${\sim}11500\,\mathrm{km}\,\mathrm{s}^{-1}$ (see Figure \ref{fig:Ejecta_composition}). This suggests the spectral formation of \ion{He}{I}\,10830 \AA~is strongly influenced by material close to the inner boundary of the He shell. This is consistent with the structure of our model as the density of He peaks close to the inner boundary of the He shell at ${\sim}13000\,\mathrm{km}\,\mathrm{s}^{-1}$ and the small mass of \nickel~ present in the He shell means the bulk of the non-thermal electrons that dominate the ionization of He likely originate from the \nickel~rich inner regions of the ejecta. The spectral formation of the \ion{He}{I}\,10830 \AA~feature being dominated by this unburnt He in the inner regions of the He shell for all epochs also explains why the feature forms at constant velocity throughout the simulation. 

In contrast, the \ion{He}{I} 10830\,\AA~feature predicted by the model of C23 appears at higher velocities and forms a separate feature to the blue wing of the \ion{Mg}{II} 10927\,\AA~feature (see C23 figure\,3). The He feature in the C23 model also evolves to lower velocities with time: at 5\,d after explosion the C23 feature is formed at ${\sim}19500\,\mathrm{km}\,\mathrm{s}^{-1}$ but forms at ${\sim}18000\,\mathrm{km}\,\mathrm{s}^{-1}$ at 2 weeks after explosion. The inner velocity of the region of unburnt He in the C23 model is at ${\sim}15000\,\mathrm{km}\,\mathrm{s}^{-1}$ (see C23 figure 1) which suggests the C23 He feature does not depend so strongly on unburnt He close to the inner boundary of the He shell. This is consistent with the distribution of unburnt He in the C23 model which peaks at ${\sim}18000\,\mathrm{km}\,\mathrm{s}^{-1}$ and shows a much more gradual drop off in density at higher velocities compared to the model presented in this study. 

Neither our model nor the model of C23 predict any clear optical \ion{He}{i} features. Additionally, while the C23 simulation did not have sufficient signal to noise to determine if a $2 \mu$m feature appears, our simulation predicts no such feature forms, despite the much stronger and more persistent \ion{He}{i} 10830\,\AA\ absorption relative to the C23 simulation. Together these models therefore suggest that \ion{He}{i} 10830\,\AA\ is the most robust spectral signature of He and for double detonation models with moderate He shell masses (such as our model and the C23 model) other \ion{He}{i} spectral lines (either optical or $2 \mu$m) that can act as corroborating signatures will not necessarily appear. As such the absence of these He features should not be used to rule out the identification of \ion{He}{i} 10830\,\AA.

Comparing the predictions of the two models also demonstrates that the spectral contribution of \ion{He}{i} 10830\,\AA\ and its evolution with phase, strongly depends on the distribution of unburnt He in the model. Given that this can show substantial differences between models and with line-of-sight, we conclude that significant variation in the role of \ion{He}{i} in shaping this feature is to be expected. This suggests there is substantial diversity in the observed \ion{He}{i} 10830\,\AA\ spectral signature throughout the SN Ia population.

\subsection{Spectral comparisons to observations}

The double detonation scenario has been discussed as a potential explanation for both low luminosity SNe~Ia, including members of the transitional sub-class, as well as normal luminosity SNe~Ia. We therefore compare our model spectra to a representative normal luminosity SN~Ia, SN 2011fe \citep{nugent2011a, pereira2013a, hsiao2013a, mazzali2014a} and a representative transitional SN~Ia, SN 2022xkq \citep{pearson2024a}. Both of these SNe have a large number of high quality early time NIR spectra and display prominent NIR spectral feature(s) at similar wavelengths to where the \ion{He}{I}\,10830 \AA~feature forms in our simulation. Therefore, comparisons between our model and the NIR features predicted by these SNe are of particular interest. To quantify if such comparisons are reasonable we first compare our model optical spectra to the optical spectra of SN 2011fe and SN 2022xkq.     

\subsubsection{Suitability of comparison objects}

The model investigated in this work has a peak absolute \textit{B}-band magnitude of -18.77, which falls between that of SN 2011fe (-19.21, \citealt{richmond2012a}) and SN 2022xkq (-18.01, \citealt{pearson2024a}). The luminosity match to each of these SNe is therefore sufficient that spectroscopic comparisons between our model and each of them is reasonable.

\begin{figure}
  \centering
  \includegraphics[width=.99\linewidth,trim={0.00 0cm 0 0},clip]{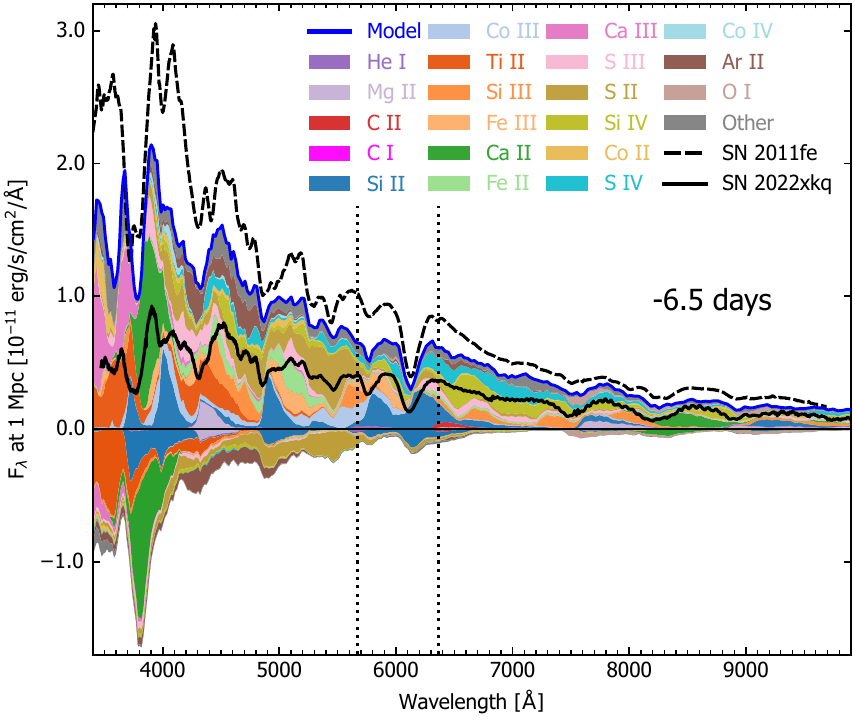}
  \includegraphics[width=.99\linewidth,trim={0.00 0.00cm 0 0},clip]{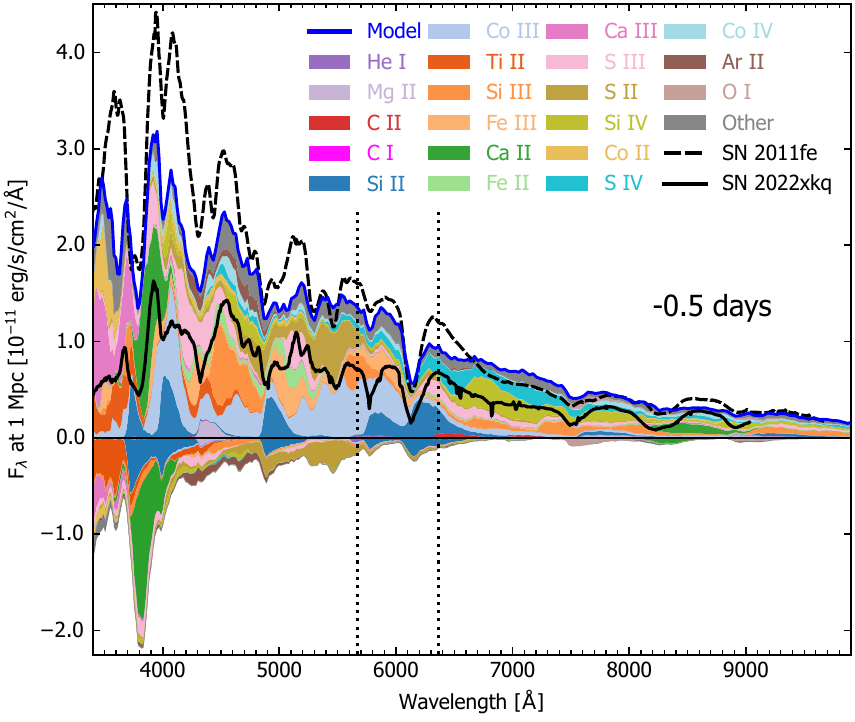}
  \includegraphics[width=.99\linewidth,trim={0.00 0.00cm 0 0},clip]{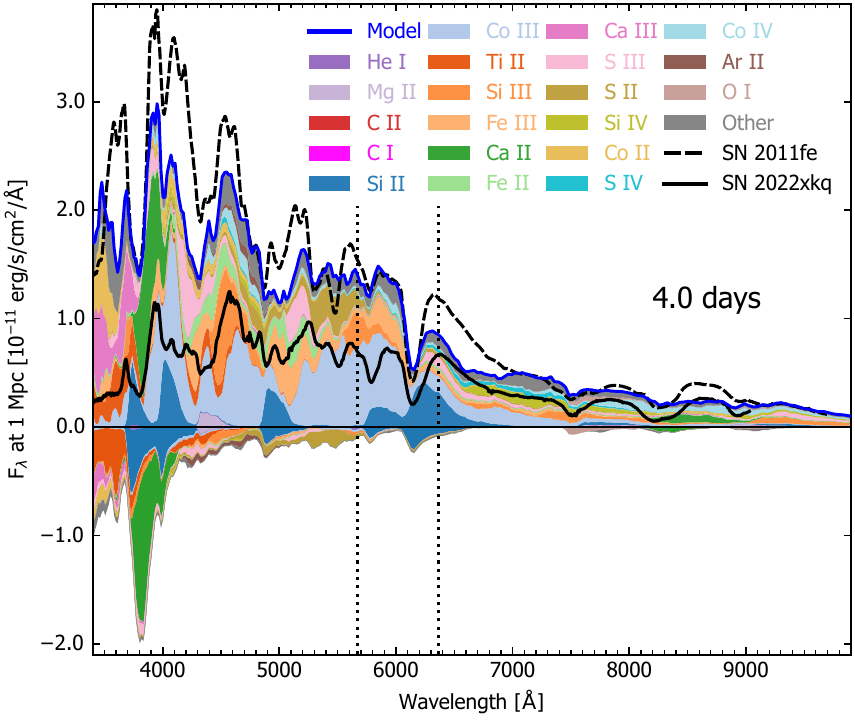}
  
  \caption{Absolute flux optical spectral comparisons between the model, with the contributions of key species indicated, SN 2011fe \citep{pereira2013a} and SN 2022xkq \citep{pearson2024a}. For reference the dotted lines indicate the wavelengths of \ion{He}{i} 5876\,\AA \ and \ion{C}{II} 6580\,\AA \ absorption. Times are relative to \textit{B}-peak.}
  \label{fig:optical_emission_absorption}
\end{figure}

We first discuss the spectral comparison between our simulation and both SN~2011fe and SN~2022xkq in the optical, to assess how well the model can account for the observed spectral features. These comparisons are shown in Figure~\ref{fig:optical_emission_absorption} where the different colours indicate the contributions of key species to the model spectra. From Figure\,\ref{fig:optical_emission_absorption} we see the spectral energy distribution (SED) predicted by our model provides a more successful match to the bluer SED exhibited by SN 2011fe (although the match to the SED of SN 2022xkq is still reasonable). This is primarily a result of both the model and SN 2011fe showing significantly less absorption blueward of ${\sim}4500$\,\AA~compared to SN 2022xkq. The model also provides a reasonable match to the strengths and velocities of a number of key features present in the spectra of both SNe, such as the prominent \ion{Si}{II} $\lambda$ 6355 and $\lambda$ 5972 features, the \ion{S}{II} $\lambda$$\lambda$ 5468 ‘‘W'', the \ion{Ca}{II} NIR triplet and the \ion{O}{I} $\lambda$ 7773 feature. However, from peak onwards, some features in the spectra of SN 2022xkq redward of ${\sim}5000$\,\AA~are noticeably stronger than those predicted by the model. In particular the \ion{Si}{II} $\lambda$ 5972, \ion{O}{I} $\lambda$ 7773 and the \ion{Ca}{II} NIR triplet are significantly stronger. The \ion{Si}{II} $\lambda$ 6355 feature does however show a relatively similar strength for the model and SN 2022xkq after peak (although the observed feature is a little too strong). We note that while the model provides a reasonable match to the \ion{Ca}{II} H\&K in both SNe, it is generally less successful at reproducing the spectral features exhibited by both SN blueward of ${\sim}5000$\,\AA. Overall, our model produces slightly better optical spectroscopic agreement with SN 2011fe over the epochs compared. However, the agreement of the model with both SNe is sufficient to justify the detailed comparisons of individual He spectral features, which are the primary focus of this study.

\label{subsec:optical_spectra}

\subsubsection{Simulated \ion{He}{i} 10830 \AA \ feature compared to observations}
\label{subsubsec:He_10830_comparisons_observations}

As discussed in Section~\ref{subsubsec:He_I_10830}, our simulation shows a strong feature due to \ion{He}{i} 10830 \AA. 
We now discuss how this feature compares to observations of the normal SN 2011fe and the transitional SN 2022xkq.

Figure\,\ref{fig:NIR_features_zoom} shows the NIR spectroscopic evolution of our model He features compared to SN 2011fe and SN 2022xkq. The full NIR spectroscopic evolution of the model compared to these SNe is shown in Figure~\ref{fig:NIR_spectral_series}. In Figures \ref{fig:NIR_features_zoom} and \ref{fig:NIR_spectral_series} the spectra have been scaled so features can be more easily compared. 

First comparing to SN 2011fe (left panel Figure~\ref{fig:NIR_features_zoom}), we see the He feature predicted by our simulation overlaps with the feature \cite{hsiao2013a} identify as \ion{Mg}{II} 10927\,\AA~in SN 2011fe. In the spectra of SN 2011fe this feature shows an extended blue wing which \cite{hsiao2013a} attribute to \ion{C}{I} 10693\,\AA~(\citealt{marion2015a} propose the same identification for SN 2014J, another normal SN~Ia). However, our model does not predict any clear feature from \ion{C}{I} 10693\,\AA~(see Section \ref{subsec:C_spectral_contribution} for further discussion). The strongest absorption of the model He feature is to the blue wing of the Mg feature in SN 2011fe. We therefore suggest this extended wing may instead be an observational signature of \ion{He}{I}\,10830 \AA. In this interpretation the observed absorption feature forms from a strong blend of \ion{Mg}{II} 10927\,\AA~and \ion{He}{I}\,10830\,\AA~with \ion{He}{I}\,10830 \AA~primarily contributing to the blue wing of the feature. We do however note the He feature predicted by our model (which represents a realisation of the double detonation scenario with a fairly substantial ejected He mass) is significantly too strong to be consistent with SN 2011fe (or SN 2014J). Therefore, further simulations for models with somewhat lower ejected He masses than the model presented here are desirable to investigate if they predict a He contribution to the blue wing of the Mg feature consistent with observations. We also note \ion{Mg}{II} 10927\,\AA~contributes to our model spectrum at wavelengths consistent with the Mg feature observed in the spectrum of SN 2011fe. However the contribution of \ion{Mg}{II} 10927\,\AA~is weaker than that of \ion{He}{I}\,10830\,\AA~in our simulation for all epochs simulated. 

Now focusing on comparisons with SN 2022xkq (middle panel of Figure \ref{fig:NIR_features_zoom}), we see SN 2022xkq also shows a strong absorption feature that \cite{pearson2024a} identify as \ion{Mg}{II} 10927\,\AA. The \ion{He}{I}\,10830 \AA~feature predicted by our model again overlaps with this \ion{Mg}{II} feature.

However, SN 2022xkq also shows a weaker secondary absorption feature to the blue of the Mg feature that disappears by peak (this feature is also observed at early times in other transitional SNe~Ia, see e.g.\,\citealt{hsiao2015a, wyatt2021a}). Our model does not predict any such double absorption feature and the \ion{He}{I}\,10830 \AA~feature predicted by our model forms at too low a velocity to match the secondary absorption feature. However, as discussed in Section\,\ref{subsec:C23_comparisons} the double detonation model of C23 does predict such a double feature with Mg \textsc{ii} 10927~\AA~forming the primary absorption feature and high velocity \ion{He}{I}\,10830\,\AA~forming the secondary feature. C23 demonstrated this simulated feature was consistent with the double absorption feature observed in the spectrum of another transitional SN, iPTF13ebh \citep{hsiao2015a} which is very similar to the feature observed in the spectrum of SN 2022xkq. We note that previous studies have suggested this secondary absorption feature observed in the spectra of some SNe~Ia is instead formed by \ion{C}{I} 10693\,\AA~(e.g.\,\citealt{hoeflich2002a, hsiao2015a, wyatt2021a, li2022a, pearson2024a}), however neither our model nor the C23 model show any such \ion{C}{I} 10693\,\AA~spectral feature (see Section \ref{subsec:C_spectral_contribution} for further discussion).

\subsubsection{Optical He spectral contribution and observations}

The simulation showed no clearly identifiable He features in the optical (as was the case for the model of C23), but nonetheless we now discuss how the optical regions with slight He contributions compare with the observations.

From Figure \ref{fig:Optical_He_spectral_series} we can see that compared to SN 2011fe and SN 2022xkq the blue wing of the \ion{Si}{ii} 5972 \AA~feature in our full model spectra predicts generally too much absorption while the model spectra in which line opacity from He is excluded does not predict enough absorption. As such, for all epochs compared, both sets of model spectra provide a similar match to the blue wing of the observed \ion{Si}{ii} 5972 \AA~feature. 

The next strongest expected He line is \ion{He}{i} 6678 \AA.
The very weak spectral contribution from \ion{He}{i} 6678 \AA \ in the simulation does not impact the agreement with either SN 2011fe or SN~2022xkq over the wavelength range to which it contributes. Therefore, the presence of strong  \ion{He}{I} 10830 \AA, as is the case for the double detonation model we present here, does not require that there are strong, clearly observable optical He features.

\subsubsection{\ion{He}{i} 2 $\mu$m feature and observations}

A clear \ion{He}{i} 2 $\mu$m feature is not predicted by this simulation. The 0.04\,\msun~of He ejected by the model we investigate here is likely towards the upper limit of the ejected He mass that can be present in a model for it to still produce reasonable agreement with the optical spectra of normal and transitional SNe~Ia \citep{kromer2010a}. The weak contribution of \ion{He}{I} 2 $\mu$m in our simulation (see Figure \ref{fig:NIR_features_zoom}) therefore suggests it is very unlikely that such a feature would be observable for any normal or transitional SNe~Ia. This is consistent with the fact that neither SN 2011fe or SN 2022xkq show any spectral feature around 2 $\mu$m. Therefore, the presence of strong \ion{He}{I} 10830 \AA\ does not require that a 2 $\mu$m feature is also observed. We note however that even if this feature can form, observing it will be challenging for ground based telescopes because the spectral contribution of the \ion{He}{I} 2 $\mu$m line predicted by the model lies very close to a telluric region (see right panel in Figure \ref{fig:NIR_features_zoom}) and thus the observed spectra are likely still impacted by the proximity to this region. For example, the observed SN 2022xkq spectrum at 0.1\,d after \textit{B}-peak shows what appears to be an absorption feature around 2 $\mu$m. However, the potential feature does not show a consistent evolution in the observed spectra and only appears for a single epoch and thus is very unlikely to be a real feature \citep{pearson2024a}. Therefore space based (e.g\, JWST) observations will likely be required to observe any potential 2 $\mu$m feature. 

\subsubsection{Spectral contribution of C  and observations}
\label{subsec:C_spectral_contribution}
\ion{C}{I} 10693 \AA~has been proposed to have an observable contribution to the blue wing of the Mg \textsc{ii} 10927~\AA~feature in SNe~Ia before peak, either as an extended blue wing (e.g.\,\citealt{hsiao2013a, marion2015a}) or a distinct secondary feature (e.g.\,\citealt{hoeflich2002a, hsiao2015a, wyatt2021a, li2022a, pearson2024a}). We therefore comment here on the spectral contribution of C in our simulation and the C23 simulation. 

Neither our model or the C23 model predicts any clear feature from \ion{C}{I} 10693 \AA. This is not unexpected as the ion population of \ion{C}{II} dominates over that of \ion{C}{I} for both simulations. This is consistent with the findings of previous studies (e.g.\,\citealt{tanaka2008a, heringer2019a}) although we do note that some simulations have shown \ion{C}{I} features can form for different classes of explosion models (e.g.\,\citealt{hoeflich2002a, blondin2013a, blondin2017a, hsiao2015a}).  

Our model does show some contribution from \ion{C}{II} in the optical. In particular, a weak \ion{C}{II} 6580 \AA~feature forms on the red shoulder of the Si \textsc{ii} 6355 \AA~feature from \tilda 7\,d before up until \textit{B}-peak. This feature appears at a velocity generally consistent with the \ion{C}{II} 6580 \AA~feature observed in the spectra of SN 2011fe and SN 2022xkq. We also see a weak spectral contribution from \ion{C}{II} 7235 \AA~(again see Figure \ref{fig:Optical_He_spectral_series}), however the contribution is so weak that it never produces any clear spectral feature. We note that while the C23 simulation also shows contributions from \ion{C}{ii} 6580 and 7235 \AA, unlike our simulation no clear \ion{C}{ii} 6580\,\AA\ feature forms due to blending with stronger lines.  

The amount of C ejected can vary significantly between different double detonation models. For example, the 3D double detonation model sequence of \cite{gronow2021a} shows a variation in C mass between models of 2.8 $\times 10^{-5}$ to 1.5 $\times 10^{-2}$\,\msun. The model we present here has an ejected C mass of 9 $\times 10^{-3}$\,\msun. While this is representative of a relatively carbon rich representation of the double detonation scenario it is still well within the range of ejected C masses predicted for double detonation models. As such the clear \ion{C}{II} 6580 \AA~feature this model predicts demonstrates that the presence of optical \ion{C}{II} can not be used to rule out the double detonation scenario. 

\section{Conclusions}
\label{sec:conclusions}
We have carried out a NLTE radiative transfer simulation including treatment for non-thermal electrons for a realisation of the double detonation scenario with a modest mass of He (0.04 \msun) that is ejected from the outer helium layer with relatively low velocities (${\sim}13000\,\mathrm{km}\,\mathrm{s}^{-1}$). The model is representative of the negative polar direction in the 3D \cite{gronow2021a} model on which it is based (where the He detonation is ignited in the positive polar direction). We find our model predicts a strong and persistent \ion{He}{I}\,10830 \AA~spectral feature that is strongly blended with the spectral contribution of \ion{Mg}{II} 10927\,\AA. This \ion{Mg}{II} feature shows an extended wing in the spectra of some normal SNe~Ia that has previously been attributed to \ion{C}{I} 10693\,\AA~\citep{hsiao2013a, marion2015a}. However, the \ion{He}{I}\,10830 \AA~feature predicted by our simulation has its strongest absorption to the blue wing of this \ion{Mg}{II} 10927\,\AA~feature and our model does not predict a \ion{C}{I} 10693\,\AA~feature forms. We therefore suggest the extended wing may instead be a spectral signature of He, although we note the He feature predicted by our particular model is too strong and persistent to be consistent with normal SNe~Ia.  

Despite the strong \ion{He}{I}\,10830 \AA~feature feature predicted by our simulation no clear optical He spectral features appear. Therefore, consistent with what C23 found for their double detonation simulation, our simulation demonstrates that the non-detection of optical He features can not be used to rule out the \ion{He}{I}\,10830 \AA~spectral feature. Our simulation also predicted no clear \ion{He}{i} 2 $\mu$m feature forms, demonstrating the absence of \ion{He}{i} 2 $\mu$m also does not rule out \ion{He}{I}\,10830 \AA\ (the C23 simulation did not have sufficient signal to noise to test the formation of the \ion{He}{i} 2 $\mu$m feature). The He mass of our model is likely towards the high end of what can be ejected while still producing reasonable agreement with the optical spectra of normal and transitional SNe~Ia \citep{kromer2010a} which suggests observed \ion{He}{I}\,10830 \AA\ spectral signatures present in the spectra of normal or transitional SNe Ia are not necessarily accompanied by additional He signatures.  

Relative to our simulation, the double detonation model presented by C23 (which also included a full treatment of NLTE and non-thermal processes) has a significantly lower He ejecta mass (0.018\,\msun), present at higher velocities (${\sim}18000\,\mathrm{km}\,\mathrm{s}^{-1}$) and is representative of the equatorial direction in the 3D \cite{gronow2020a} model on which it is based. The contribution of \ion{He}{I}\,10830 \AA\ is weaker and present at higher velocities for the C23 simulation, forming a distinct feature blueward of \ion{Mg}{II} 10927\,\AA\ which is no longer clearly present by peak. This feature is consistent with a feature present in the pre-peak spectra of observed transitional SNe Ia \citep{hsiao2015a, wyatt2021a, pearson2024a}. Together these simulations demonstrate that the He signatures predicted by double detonation models can show substantial variation depending on the exact realisation of the ejecta structure, and in particular the distribution of unburnt He, which is both model and line-of-sight dependent. This implies significant diversity is expected in the \ion{He}{I}\,10830 \AA\ spectral signature across the observed SN Ia population. This should be tested with observations of the pre-peak NIR spectroscopic evolution for a larger sample of SNe Ia. 

Future work should also focus on full NLTE multi-dimensional radiative transfer simulations of double detonation models that will allow the viewing angle variation of predicted He spectral signatures to be explored. In particular, since both our simulation and the C23 simulation predict a \ion{He}{I}\,10830 \AA\ spectral signature too strong compared to normal SNe Ia, double detonation models with lower masses of ejected He should be prioritised. Such simulations will be key to determining whether realisations of the double detonation scenario exist that predict He spectral signatures consistent with the NIR spectra of normal SNe~Ia and will also allow us to probe the lower limit of He that can be ejected while still producing an observable signature.

\section*{Acknowledgements}
FPC and SAS, acknowledge funding from STFC grant ST/X00094X/1. This work used the DiRAC Memory Intensive service (Cosma8) at Durham University, managed by the Institute for Computational Cosmology on behalf of the STFC DiRAC HPC Facility (www.dirac.ac.uk). The DiRAC service at Durham was funded by BEIS, UKRI and STFC capital funding, Durham University and STFC operations grants. Access to DiRAC resources was granted through a Director’s Discretionary Time allocation in 2023/24. This work also used the DiRAC Data Intensive service (CSD3) at the University of Cambridge, managed by the University of Cambridge University Information Services on behalf of the STFC DiRAC HPC Facility (www.dirac.ac.uk). The DiRAC component of CSD3 at Cambridge was funded by BEIS, UKRI and STFC capital funding and STFC operations grants. DiRAC is part of the UKRI Digital Research Infrastructure. The authors gratefully acknowledge the Gauss Centre for Supercomputing e.V. (www.gauss-centre.eu) for funding this project by providing computing time on the GCS Supercomputer JUWELS at Jülich Supercomputing Centre (JSC).
CEC acknowledges funding by the European Union (ERC, HEAVYMETAL, 101071865). LJS acknowledges support by the European Research Council (ERC) under the European Union’s Horizon 2020 research and innovation program (ERC Advanced Grant KILONOVA No. 885281).
LJS acknowledges support by Deutsche Forschungsgemeinschaft (DFG, German Research Foundation) - Project-ID 279384907 - SFB 1245 and MA 4248/3-1. JMP acknowledges the support of the Department for Economy (DfE). The work of FKR is supported by the EuropeanUnion (ERC, ExCEED, 101096243) and by the Deutsche Forschungsgemeinschaft (DFG, German Research Foundation) - project number 537700965. Views and opinions expressed are however those of the author(s) only and do not necessarily reflect those of the European Union or the European Research Council. Neither the European Union nor the granting authority can be held responsible for them. This work was supported in part by the European Union (ChETEC-INFRA, project no. 101008324), NSF/IReNA and the Klaus Tschira Foundation. NumPy and SciPy \citep{oliphant2007a}, Matplotlib \citep{hunter2007a}  and \href{https://zenodo.org/records/8302355} {\textsc{artistools}}\footnote{\href{https://github.com/artis-mcrt/artistools/}{https://github.com/artis-mcrt/artistools/}} \citep{artistools2025a} were used for data processing and plotting.

\section*{Data Availability} 
The spectra presented here will be made available on the Heidelberg
supernova model archive HESMA\footnote{\href{https://hesma.h-its.org}{https://hesma.h-its.org}}~\citep{kromer2017a}.



\bibliographystyle{mnras}
\bibliography{references} 




\appendix

\section{NIR spectral evolution}
\begin{figure*}
    \centering
    \begin{subfigure}[t]{0.495\textwidth}
        \centering
	\includegraphics[width=1\linewidth,trim={2.7cm 2.0cm 2.1cm 2.5cm},clip]{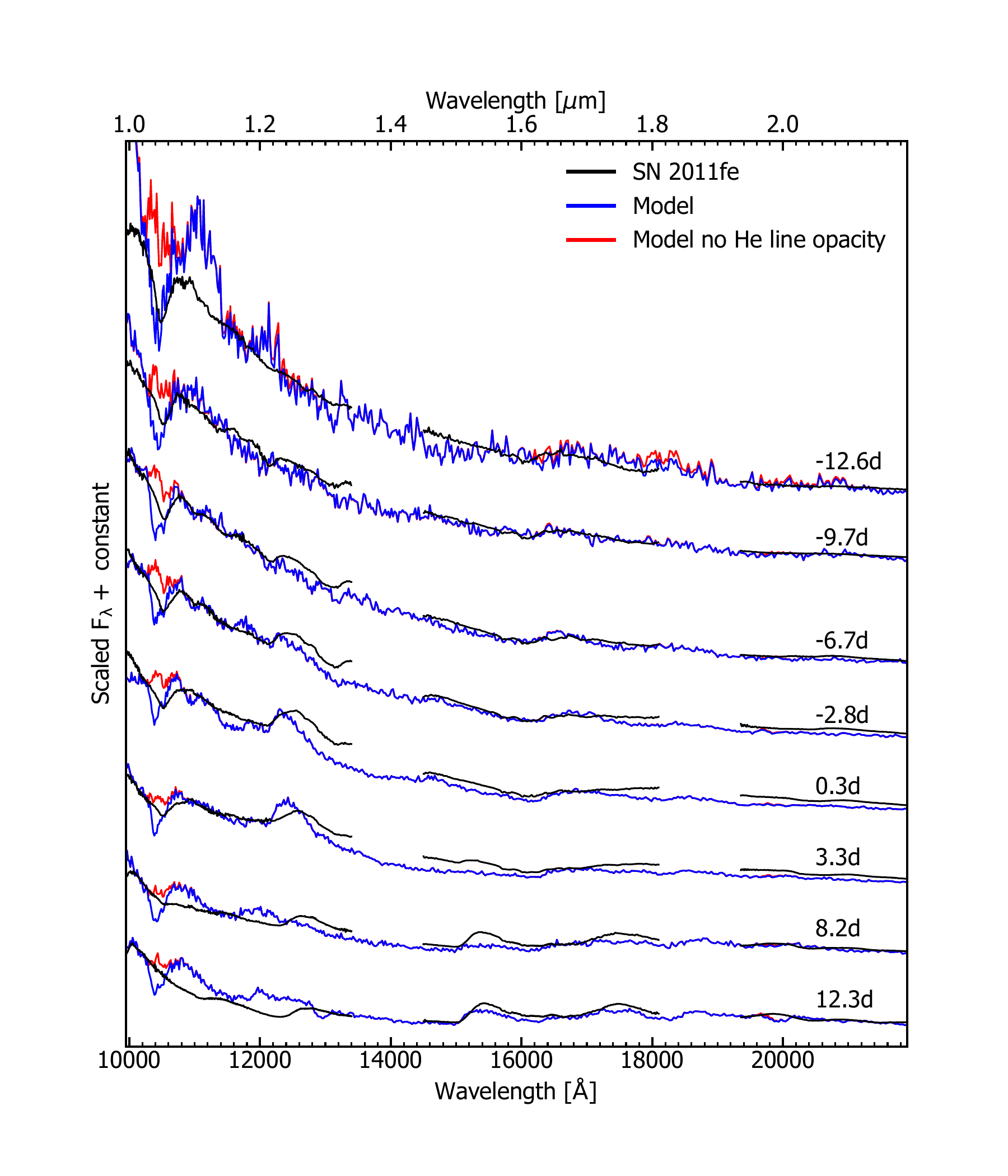} 
    \end{subfigure}
    \begin{subfigure}[t]{0.495\textwidth}
        \centering
	\includegraphics[width=1\linewidth,trim={2.7cm 2.0cm 2.1cm 2.5cm},clip]{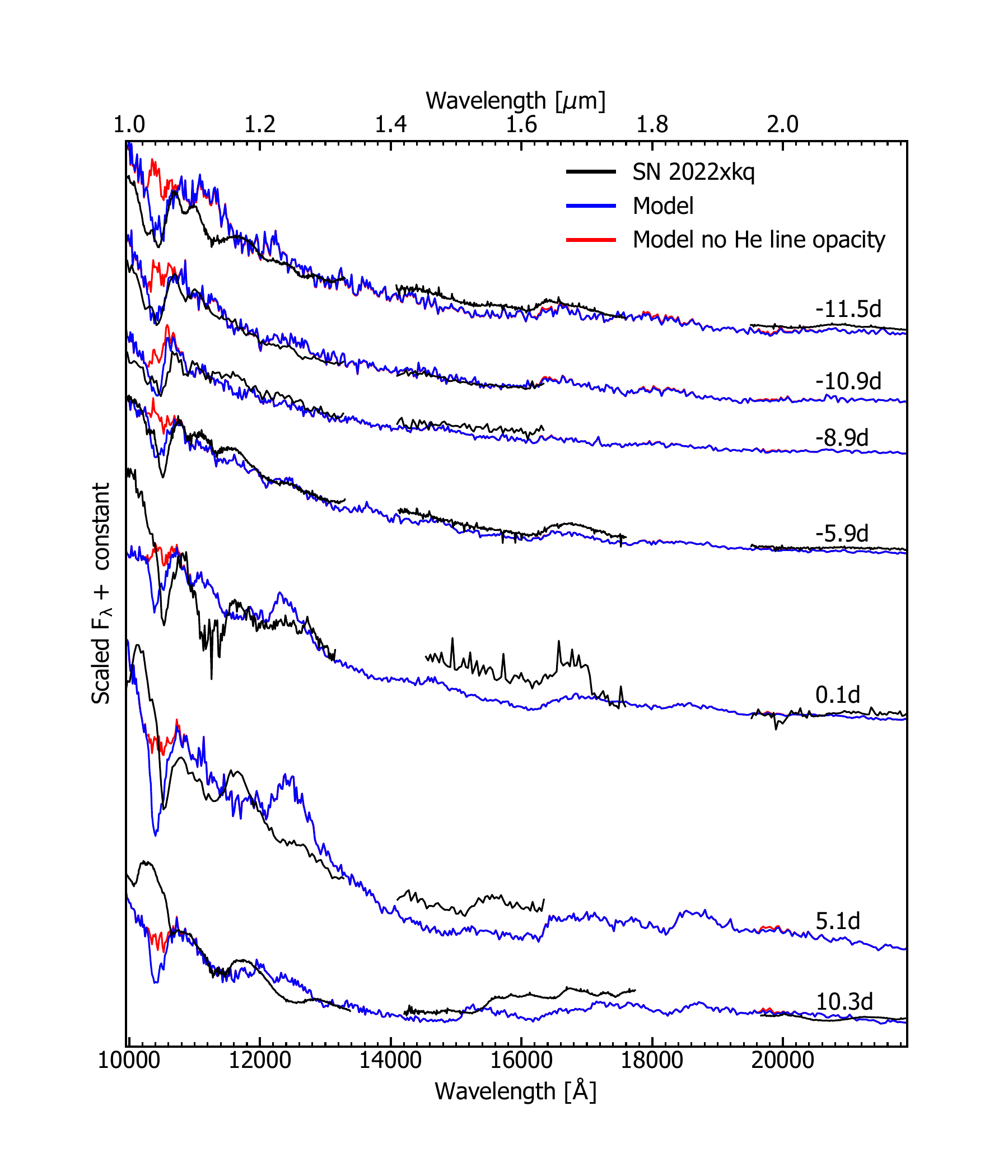}    
    \end{subfigure}
    \caption{NIR spectroscopic evolution of our full model spectra and our model spectra in which the line opacity for He is not included, compared to SN 2011fe (\citealt{hsiao2013a}, left) and SN 2022xkq (\citealt{pearson2024a}, right). Times are relative to \textit{B}-peak. Note, that for clarity the spectra of SN 2022xkq have been rebinned and the regions impacted by tellurics have been removed.}
    \label{fig:NIR_spectral_series}
\end{figure*}

In Figure \ref{fig:NIR_spectral_series} we show the NIR spectra for our full model and a model in which the line opacity of He is omitted. These are compared to observations of SN 2011fe \citep{hsiao2013a} and SN 2022xkq \citep{pearson2024a}.  

\bsp	
\label{lastpage}
\end{document}